\documentclass[12pt]{article}
\usepackage[left=2.4cm,right=2.4cm,top=2.8cm,bottom=2.8cm]{geometry}
\usepackage[colorlinks=true,linkcolor=black,citecolor=black,urlcolor=blue,filecolor=black]{hyperref}

\usepackage{amsmath}
\usepackage{amsfonts}
\usepackage{amssymb}
\usepackage[usenames,table]{xcolor}
\usepackage{dsfont}
\usepackage{mathrsfs}

\usepackage{verbatim}

\begin{document}
\vspace{30pt}

\begin{center}


{\Large{\sc The complete action for $\mathcal{N}=2$ de Sitter 
pure supergravity}\\[10pt]}

\vspace{-5pt}
\par\noindent\rule{350pt}{0.4pt}


\vspace{20pt}

{\sc 
Nicolas Boulanger, Vasileios A. Letsios and Sylvain Thom\'ee}

\vspace{35pt}
{\it\small
Service de Physique de l'Univers, Champs et Gravitation,\\
Universit{\'e} de Mons -- UMONS,
20 place du Parc, 7000 Mons, Belgium}
\vspace{10pt}

{\tt\small 
\href{mailto:nicolas.boulanger@umons.ac.be}{nicolas.boulanger@umons.ac.be},
\href{mailto:vasileios.letsios@umons.ac.be}{vasileios.letsios@umons.ac.be},
\href{mailto:sylvain.thomee@umons.ac.be}{sylvain.thomee@umons.ac.be}
}

\vspace{40pt} {\sc\large Abstract} \end{center}

\noindent

Supergravity theories in de Sitter spacetime are known to be very constrained, 
and rather unnatural within String/M Theory. 
We revisit the seminal paper by Pilch, van Nieuwenhuizen and Sohnius,
where the possible existence of a real Lagrangian for ${\cal N}=2$ 
pure supergravity in four-dimensional de Sitter spacetime was pointed out. 
We clarify several issues related to the non-unitarity of the theory 
and explicitly construct the unique, complete theory searched for long 
ago by the aforementioned authors. We argue that the lack of unitarity 
of the Lorentzian theory may be revisited in the Euclidean approach to 
de Sitter quantum gravity, where alternative definitions of unitarity 
can be introduced.
\newpage


\section{Introduction}

On the one hand, both the primordial and current Universe seem to be well 
approximated by de Sitter geometry. On the other hand, the 
non-renormalisability of Einstein's theory of gravity calls for an 
extension of it that should behave better in the high-energy limit. 
Supergravities were born in the late seventies 
\cite{Deser:1976eh, Freedman:1976xh} with the 
hope to give a renormalisable model of quantum field theory containing
Einstein's theory as a consistent truncation.
This hope was later shown to be unrealistic, string theory 
offering more promises to attain UV finiteness while at the same time 
containing a massless spin-2 mode associated with the graviton, 
see e.g. \cite{Green:1987sp}.
More recently, however,  
the works \cite{Bern:2006kd,Bern:2007hh, Bern:2008pv, Bern:2009kd, Bern:2014sna,Bern:2018jmv} and references therein prompted a renewed 
interest  in supergravity theories.
The very first paper where supergravity was considered in a 
de Sitter background is \cite{Pilch:1984aw} by Pilch,  
van Nieuwenhuizen, and Sohnius.
At that time, when supergravity theories were born, our Universe 
was believed to have no cosmological constant, and the paper 
\cite{Pilch:1984aw} appeared to be of more academic relevance 
than phenomenological necessity.
Still, this paper is seminal in that, 
for the first time, it clearly described the problems encountered 
in constructing supergravity theories in de Sitter (dS$_4$) spacetime 
due to a clash between the reality conditions and the necessary gauge 
invariance to be imposed on a vector-spinor to only propagate the 
helicity degrees of freedom $\pm 3/2$. Supergravity favours 
a negative cosmological constant \cite{Townsend:1977qa} 
--- although this was not stressed in the seminal work \cite{Townsend:1977qa},
which also adopted misleading terminology in use at the time,
when the maximally symmetric spacetime of negative constant 
curvature was termed ``de Sitter'' instead of ``anti-de Sitter'' ---, 
therefore it was crucial that the paper \cite{Pilch:1984aw} pointed out 
the possible existence of a consistent supergravity theory in 
de Sitter spacetime at the price of a ghostly graviphoton,
also showing that the minimal number of supersymmetries in 
dS$_4$ supergravity is ${\cal N}=2\,$. 

Since then, it was discovered \cite{SupernovaCosmologyProject:1998vns, SDSS,PlanckCollab} that our Universe possesses a 
small,  positive cosmological constant, de Sitter spacetime 
being the simplest and most symmetric solution to Einstein's equations 
with a positive cosmological constant. 
String/M theory, although remarkable in that it is a quantum 
theory that contains gravity, does not seem to naturally allow
for de Sitter vacua, see e.g. \cite{Danielsson:2018ztv,Maldacena:2000mw,Lust:2022lfc}, 
but see also \cite{Kachru:2003aw,Cribiori:2019hrb,Burgess:2024jkx} for some positive results in this direction.
Independently of the issue of de Sitter vacuum 
in string theory, the study of quantum field theories in de Sitter 
spacetime is a very important topic, dating back to the sixties, 
see for example 
\cite{Geheniau:1966,Schomblond:1976xc,Bunch:1978yq,Higuchi:1986py,Bros:1994dn,Deser:2003gw} 
for a very small set of references.
The field has witnessed a recent surge of interest, see e.g. 
\cite{Anninos:2017eib,Anninos:2020hfj,Letsios-non-uni-I,Melville:2024ove,Du:2025zpm}
and references therein.

In the present paper, we revisit the work \cite{Pilch:1984aw} and 
complete it in several directions. First, we rigorously prove the existence 
and uniqueness of the simplest possible supergravity theory in 
four-dimensional de Sitter (dS$_4$) spacetime, using tools 
\cite{Barnich:1993vg,Henneaux:1997bm} unavailable at the time of 
\cite{Pilch:1984aw}. The latter work gives an action invariant under 
gauge supersymmetry, only up to terms quartic in fermions, 
while in the present paper we provide the unique, complete Lagrangian 
that is invariant under ${\cal N}=2$ local supersymmetry, to all orders 
in the coupling constant (Newton's constant), and in particular, to all
orders in the fermionic fields. 
Second, we provide insight concerning the issue of non-unitarity 
discussed in \cite{Pilch:1984aw}. In particular, although we confirm that 
the graviphoton necessarily has the wrong sign in front of its kinetic term, 
we explain that the massless gravitino also has unitarity issues in de 
Sitter, not observed at the time of \cite{Pilch:1984aw}.
Using recent results \cite{Letsios-non-uni-I, Letsios-non-uni-II, Higuchi:2025pbc} showing that the quantisation of the free Dirac gravitino 
on Lorentzian dS$_{4}$ leads to indefinite-norm states, we clarify that the graviphoton 
is not the only source of non-unitarity for $\mathcal{N}=2$ de Sitter 
supergravity, contrary to the standard lore.

We conclude this paper by discussing the relevance of our results in the 
context of the Euclidean approach to quantum gravity with positive cosmological 
constant, suggesting future directions. The Euclidean approach to de Sitter 
quantum gravity enjoys renewed interest thanks to invigorating recent 
results \cite{Anninos:sphere, Maldacena:2024spf, Anninos:2025mje}, such as  
1-loop expressions for the sphere path integral in terms of group-theoretic 
characters of the de Sitter group. 
We explain how our results for the existence and uniqueness of 
pure $\mathcal{N}=2$ de Sitter supergravity in Lorentzian signature 
set the basis for the study of the Euclidean version of the theory. Motivated by recent findings for the 1-loop sphere partition function for the gravitino \cite{Anninos:2025mje}, we explain why the unitarity issues of the Lorentzian supergravity theory may turn out to be 
harmless in Euclidean signature, suggesting promising future directions. 
In particular, this invites us to emphasise the importance of addressing a 
particular open question; the question of consistency 
of pure $\mathcal{N}=2$ de Sitter supergravity in Euclidean signature, see also discussions in \cite{Anninos:2025mje}.

Let us now review what is known for the two  kinds of obstacles to the existence of 
unitary, unbroken de Sitter supersymmetry.\footnote{For references 
about broken $\mathcal{N}=1$ supersymmetry and de Sitter, 
see e.g. \cite{Freedman:1976uk,Lindstrom:1979kq,Bergshoeff:2015tra,Bergshoeff:2016psz,Bandos:2016xyu,Antoniadis:2020qoj}.} 

\begin{enumerate}
    \item \textbf{Non-unitarity of global supersymmetry on a fixed de Sitter background.}  This problem stems from abstract representation theory of the de Sitter superalgebra \cite{Pilch:1984aw,Lukierski:1984it}. The de Sitter superalgebra is the supersymmetric extension of the dS algebra $so(4,1)$, i.e. apart from the ten Grassmann-even $so(4,1)$ generators, there also exist spinorial supercharges ${Q}^{(i)}_{A}$ (Grassmann-odd generators). The index $A$ is a spinor index referring to the fundamental spinor representation of $so(4,1)$, and $i$ is an extended supersymmetry index. Let us assume that the spinorial generators are Dirac spinors for simplicity - however, the analysis presented in this paragraph holds also for the case where a symplectic Majorana reality condition is introduced for the supercharges \cite{Pilch:1984aw, Lukierski:1984it}. The main feature of the dS superalgebra is that the following trace vanishes 
    \begin{align}\label{non-unitarity equation susy algebra}
      \sum_{A,i} \{ {Q}^{(i)}_{A}, {Q}^{(i)A \, \dagger}   \} = 0  \;.
    \end{align}
    This is physically justified by the absence of a generator of $so(4,1)$ with bounded spectrum that could be placed on the right-hand side of (\ref{non-unitarity equation susy algebra}).
    The vanishing of the trace in (\ref{non-unitarity equation susy algebra}) implies that, if one tries to realise these spinorial supercharges as operators acting on a Hilbert space, then  all non-trivial representations of the dS superalgebra must have indefinite norm.  Similar unitarity no-go theorems exist for the super-extensions of $so(3,1)$ and $so(5,1)$ \cite{Lukierski:1984it}. Interestingly, unitary representations exist for the  super-extension of $so(2,1)$ \cite{Lukierski:1984it}. See \cite{Anninos:2023exn} for a recent discussion of the Euclidean approach to 2-dimensional supergravity with positive cosmological constant, where unitarity obstacles can be avoided.

    \item \textbf{Non-unitarity of dS$_{4}$ pure supergravity and lack 
    of a full theory.} 
    As mentioned earlier, a local and real action functional for $\mathcal{N}=2$ supergravity 
    with positive cosmological constant, which is invariant under local supersymmetry, has 
    been proposed in \cite{Pilch:1984aw}, where the terms quartic in 
    fermion were \textbf{not} given. The dynamical fields of this 
    theory are:  a real vierbein, a real (gravi)photon, and two symplectic 
    Majorana gravitini. The double number of symplectic Majorana  gravitini  is 
    related to the fact that  a conventional Majorana condition is not available, 
    see e.g. \cite{Pilch:1984aw, Higuchi:2025pbc}. More specifically, 
    this difficulty stems from the fact that the supercharges of the de Sitter 
    superalgebra must behave as spinors of $Spin(4,1)$, and there is no Majorana 
    spinor representation for this group. Nevertheless, a symplectic Majorana 
    condition can still be used. We stress that the terms that are quartic in 
    gravitini were not provided in \cite{Pilch:1984aw}, i.e. the local supersymmetry 
    invariance of the action was checked up to terms quartic in fermions.  The main 
    result of Ref.~\cite{Pilch:1984aw} was that, although the $\mathcal{N}=2$  
    supergravity action is invariant under local supersymmetry (up to terms quartic 
    in fermions), the  photon kinetic 
    term has the wrong sign, leading to the presence of negative norms. Thus, if 
    a full theory exists, then it will have negative-norm states. However, the 
    questions of whether  a full theory of $\mathcal{N}=2$ supergravity with 
    positive cosmological constant exists, and whether the theory is unique, were 
    left open in \cite{Pilch:1984aw}. The present paper addresses these questions.
    \end{enumerate}
   
\section{The free theory}

Consider a free model on a dS$_4$ background with metric $\bar{g}_{\mu\nu}$ and cosmological constant $\Lambda=3\lambda^2$ containing: a Dirac massless spin-3/2 vector-spinor field $\phi_\mu$ [spinor indices are suppressed], a real massless spin-2 symmetric tensor field $h_{\mu\nu}$, and a real massless spin-1 field $A_{\mu}$. The action reads
\begin{equation}
\begin{aligned}
    S_0  = \int d^4x \sqrt{-\bar{g}} (&- \bar{\phi}_\mu \gamma^5 \gamma^{\mu \nu \rho}\Phi_{\nu\rho} -\sigma \frac{1}{4} F^{\mu \nu} F_{\mu \nu} \\
    &-\frac{1}{2} \nabla^\rho h^{\mu \nu}  \nabla_\rho h_{\mu \nu} + \nabla_\rho h^{\mu \nu} \nabla_\mu h_\nu{}^\rho - \nabla_\mu h \nabla_\nu h^{\mu \nu} \\
    &+ \frac{1}{2} \nabla^\mu h \nabla_\mu h + 3\lambda^2 h_{\mu\nu}h^{\mu \nu}- \frac{3}{2} \lambda^2 h^2)\;,\label{eq:ActionS0}
\end{aligned}
\end{equation}
where $h = h^{\mu}\,_{\mu}$, $\Phi_{\nu\rho}=\nabla_{[\nu} \phi_{\rho]} +  \frac{i}{2}\lambda \gamma_{[\nu}\phi_{\rho]}$ and $F_{\mu\nu}=\nabla_\mu A_\nu-\nabla_\nu A_\mu$ \footnote{The symbol $\nabla_{\mu}$ denotes the Lorentz-covariant derivative of the background de Sitter geometry, where we use the convention $[\nabla_\mu,\nabla_\nu] \psi =\frac{1}{2}\lambda^2\gamma_{\mu\nu}\psi$, with $\psi$ a four component spinor. We follow the strength-one (anti)symmetrisation convention, with (square) round brackets indicating (anti)symmetrisation of the enclosed indices. For example, $2\nabla_{[\nu} \phi_{\rho]}=\nabla_{\nu} \phi_{\rho}-\nabla_{\rho} \phi_{\nu}$ and $2\nabla_{(\mu}\xi_{\nu)}=\nabla_{\mu}\xi_{\nu}+\nabla_{\nu}\xi_{\mu}$.}. The parameter $\sigma \in \{ +1,-1\}$ allows one to choose the sign in front of the free spin-1 Lagrangian. The value $\sigma=+1$ corresponds to the canonical sign of a unitary theory, while if $\sigma=-1$, the spin-1 sector will include negative-norm states. The free gauge transformations that leave $S_0$ invariant, i.e. $\delta_0S_0=0$, are 
\begin{align}
    \delta_0 h_{\mu \nu} &=2 \nabla_{(\mu} \xi_{\nu)}\,, \quad \delta_0 A_\mu = \nabla_\mu \pi\;,\\
    \delta_0 \phi_\mu &= \nabla_\mu \epsilon + \frac{1}{2} i \lambda \gamma_\mu \epsilon\,, \quad
    \delta_0 \bar{\phi}_\mu = \nabla_\mu \bar{\epsilon} + \frac{1}{2} i \lambda \bar{\epsilon}\gamma_\mu\,, \label{eq:FreeGaugeTfbarPhi} 
\end{align}
where $\xi_{\nu}$ is a vector gauge parameter, $\epsilon$ is a spinor gauge parameter and $\pi$ is a scalar one.

The equations of motion derived from the free Dirac gravitino Lagrangian in \eqref{eq:ActionS0} describe the degrees of freedom of a gravitino gauge potential on a fixed de Sitter background. The physical mode solutions of these equations on global dS$_{4}$ have been discussed in detail in \cite{Letsios-non-uni-I, Letsios-non-uni-II, Higuchi:2025pbc}, and it has been shown that they furnish a direct sum of fermionic discrete series unitary irreducible representations (UIRs) of $so(4,1)$. A curious field-theoretic feature of these representations is that physical modes of helicity $-3/2$ have a different positive-definite, $so(4,1)$-invariant inner product compared to 
the helicity $+3/2$ physical modes. In fact, the two inner products differ 
by a factor of $-1$. At the level of mode solutions, a different inner product 
can be used for each fixed-helicity sector (each fixed-helicity sector furnishes 
independently a discrete series UIR). However, from a quantum field theory viewpoint, the $so(4,1)$-invariant inner product must be the same for both helicities, 
if we insist on the locality of the theory \cite{Higuchi:2025pbc}. 
Thus, the free-theory Dirac spin-3/2 Lagrangian used in \eqref{eq:ActionS0} 
gives rise to indefinite norms upon quantisation. At the free level, there are 
two ways to achieve the unitary quantisation of the gravitino field on dS$_4$. 
The first approach is to impose the on-shell anti-self-duality constraint 
to the gravitino field strength, which removes all negative-norm states, 
and thus quantise the chiral gravitino theory \cite{Higuchi:2025pbc}. 
The second approach is to use an operator algebra perspective as a guiding principle, which leads to 
positive-definite inner product for the mode solutions that cannot 
be derived from a local Lagrangian \cite{Anninos:2025mje}.

We have shown that the gravitino Lagrangian in \eqref{eq:ActionS0}  
is equivalent to the one used by Pilch, van Nieuwenhuizen, and Sohnius in 
\cite{Pilch:1984aw}, by properly expanding the Dirac spin-3/2 gravitino 
into a pair of symplectic Majorana gravitini, and performing a similarity 
transformation of the gamma matrices. We conclude that the gravitino sector of
\cite{Pilch:1984aw, Sohnius:1985ne} has  indefinite norm states, i.e. the 
graviphoton kinetic term is not the only source of non-unitarity, 
which was not noticed in these two papers.

\section{Consistent interactions at cubic order}

Starting from the free theory \eqref{eq:ActionS0}, 
we classified all the possible local, cubic interactions among the 
various fields of the spectrum. We used the cohomological 
reformulation of the Noether method given in 
\cite{Barnich:1993vg,Henneaux:1994lbw}, 
based on the antifield formalism of  
\cite{Becchi:1975nq,Tyutin:1975qk,Batalin:1981jr,Batalin:1983ggl,BatalinErratum} 
and found two non-Abelian cubic vertices in the fermionic sector. 
First, there is the gravitational coupling of the Dirac spin-3/2 field, 
denoted  $L_1^{(\phi-h)}$, with coupling constant $\alpha_1$. 
Second, we recover the electromagnetic coupling 
of the Dirac spin-3/2 field, denoted $L_1^{(\phi-A)}$, with coupling 
constant $\alpha_2$.
These vertices, as well as the corresponding deformations to the gauge 
transformations, are given in Appendices \ref{App: coupling spin-3/2 with spin-2} 
and \ref{App: coupling spin-3/2 with spin-1}. 

In the bosonic sector, there are two cubic vertices, 
the non-Abelian Einstein-Hilbert self-coupling $L_1^{EH}$, with coupling 
$\kappa$, and the Abelian gravitational coupling of the Maxwell field
$L_1^{(h-A)}$, with coupling constant $\alpha_3$. 
Both of these couplings, as well as the 
corresponding deformations of the gauge transformations and of the gauge algebra, 
are given in e.g. \cite{Boulanger:2024hrb} for both de Sitter and anti-de Sitter 
backgrounds.

We have shown that the Jacobi identities for the gauge algebra
impose the following necessary conditions on the coupling constants:
\begin{equation}
     2\,\sigma\,\alpha_2^2 + \alpha_1^2=0\;,  
     \quad\alpha_1=- \frac{\kappa}{2}\;. \label{eq:QuarticOrderRelations}
\end{equation}
As $\alpha_2$ and $\alpha_1$ are real constants, ensuring the reality of the deformed action $S$, the parameter $\sigma$ has to 
take the value $-1$. 
This leads to an opposite relative sign between the spin-1 and spin-2 kinetic 
terms, see eq. \eqref{eq:ActionS0}. 
Thus, we confirm the conclusions of \cite{Pilch:1984aw} concerning the ghostly 
nature of the graviphoton, following a different path.
Note that the derivation of eq. \eqref{eq:QuarticOrderRelations} does not require 
the explicit knowledge of the form of quartic vertices and of the corresponding 
gauge transformations.

\section{The complete theory}

\subsection{Superalgebra, fields and gauge parameters}

We denote $M_{ab}$ and $P_a$ $(a,b=0,1,2,3)$ the generators 
of $SO(4,1)$, the isometry group of 4-dimensional de Sitter spacetime. 
We introduce a supersymmetry charge, $Q$, which is a Dirac spinor, 
its conjugate $\bar{Q}$ and a $U(1)$ generator $T$.
The non-vanishing commutators and anti-commutators are
\begin{align}
    [M_{ab},M_{cd}]& = 2 \eta_{a[c} M_{d]b} -  2 \eta_{b[c} M_{d]a}\;, \label{eq:So(14)}
    \\\quad[P_{a},M_{bc}]&= 2 \eta_{a[b} P_{c]}\;,\quad
     [P_a, P_b]=-\lambda^2M_{ab}\;, \nonumber\\
    [T, Q]&= i \frac{\lambda\kappa}{\sqrt{2}} Q\;,\quad [T, \bar{Q}]=-i\frac{\lambda\kappa}{\sqrt{2}}\bar{Q}\;,\\
    [M_{ab}, Q]&=-\frac{1}{2} \gamma_{ab}Q\;,\quad[M_{ab}, \bar{Q}]=\frac{1}{2} \bar{Q}\gamma_{ab}\;,\\
    [P_a, Q]&= \frac{i\lambda}{2} \gamma_a Q\;,\quad [P_a, \bar{Q}]=\frac{i\lambda}{2} \bar{Q} \gamma_a\;,\\
    \{Q, \bar{Q}\}&=-\frac{1}{4} \gamma^5\gamma^aP_a + \frac{i \lambda}{8} \gamma^5\gamma^{ab} M_{ab} - \frac{ 1}{2\sqrt{2} \kappa}\gamma^5 T\;,\label{eq:AntiCommQQbar}    
\end{align}
where the bosonic generators are taken to be anti-hermitian.
We checked that the (super-)Jacobi identities are satisfied, 
and thus, the generators in \eqref{eq:So(14)}-\eqref{eq:AntiCommQQbar} 
define a $\mathcal{N}=2$ de Sitter Lie superalgebra, with bosonic 
subalgebra isomorphic to $so(4,1)\oplus u(1)$. 
The de Sitter superalgebras have been classified in 
\cite{Parker:1980af, Lukierski:1981ht, Lukierski:1984it, Pilch:1984aw}.

One introduces a Grassmann-even gauge field $W_\mu$ and a Grassmann-even gauge 
parameter $\varepsilon$ that are both valued in the adjoint representation of the 
superalgebra:
\begin{equation}
    W_\mu := e_\mu^aP_a + \tfrac{1}{2}\omega_\mu{}^{ab}M_{ab} 
    + \bar{\psi}_\mu \,Q + \bar{Q}\, \psi_{\mu} + A_\mu T\;,
\end{equation}
\begin{equation}
    \varepsilon:=\varepsilon^a_P P_a + \frac{1}{2} \varepsilon_M^{ab} M_{ab} + \bar{\varepsilon}_Q \,Q +\bar{Q}\, \varepsilon_{Q}+ \varepsilon_T \,T\;.
\end{equation}
The transformation of the gauge field $W_\mu$ under the gauge group is 
defined as $\delta^G\, W_\mu := \partial_\mu \varepsilon + [W_\mu, \varepsilon]\,.$
The gauge field component $e_\mu^a$ is assumed to be invertible, with inverse 
denoted $e^\mu_a$, and $e_\mu^ae^\nu_a=\delta^\nu_\mu$. 
As usual, $g_{\mu\nu}:=e_\mu^a \eta_{ab}e_\nu^b$. 
Following the gauge theory prescription, we introduce the curvatures 
of the gauge group 
\begin{equation}
    \mathcal{R}_{\mu\nu} := 2\partial_{[\mu} W_{\nu]} +[W_\mu,W_{\nu}]\;,\label{eq:CurvGroupR}
\end{equation}
\begin{equation}
\begin{aligned}
    \mathcal{R}_{\mu\nu}=& \,\mathcal{R}_{\mu\nu}^a(P)\,P_a
    +\tfrac{1}{2}\mathcal{R}_{\mu\nu}{}^{ab}(M)M_{ab}+ \bar{\mathcal{R}}_{\mu\nu}(Q)\,Q + \bar{Q}\,\mathcal{R}_{\mu\nu }(Q)+ \mathcal{R}_{\mu\nu}(T)T\;.
\end{aligned}
\end{equation}
Explicitly, the Lorentz curvature $\mathcal{R}_{\mu\nu}{}^{ab}(M)$ reads
\begin{equation}
    \mathcal{R}_{\mu\nu}{}^{ab}(M)=R_{\mu\nu}{}^{ab}-2 \lambda^2 e_{[\mu}^{a} e_{\nu]}^{b} +\frac{i \lambda}{2} \bar{\psi}_{[\mu} \gamma^5 \gamma^{ab} \psi_{\nu]}\;, \label{eq:ExpansionR(M)}
\end{equation}
where $R_{\mu\nu}{}^{ab}=2 \partial_{[\mu} \omega_{\nu]}{}^{ab} + 2 \omega_{[\mu}{}^{cb} \omega_{\nu]}{}_c{}^{a}$ is the Riemann tensor.
Also, $\mathcal{R}_{\mu\nu}(T)=F_{\mu\nu}-\frac{1}{\sqrt{2}\kappa}\bar{\psi}_{[\mu}\gamma^5\psi_{\nu]}\;$.

\subsection{The action}
Inspired by the construction of $\mathcal{N}=2$ supergravity in anti-de Sitter 
spacetime \cite{Fradkin:1976xz, Freedman:1976aw} as presented in 
\cite{MacDowell:1977jt, Townsend:1977fz}, we propose the following action of 
$\mathcal{N}=2$ de Sitter supergravity:
\begin{equation}
\begin{aligned}
    I[e,\omega,\bar{\psi}, \psi,A] = &\int d^4x \Big(\; \frac{1}{32 \kappa^2 \lambda^2}\mathcal{R}_{\mu\nu}{}^{ab}(M) \mathcal{R}_{\rho \sigma}{}^{cd}(M) \varepsilon^{\mu\nu\rho\sigma}\varepsilon_{abcd}\\ 
    &-\frac{1}{8 \kappa^2 \lambda} \bar{\mathcal{R}}_{\mu\nu}(Q) \mathcal{R}_{\rho\sigma}(Q) \varepsilon^{\mu\nu\rho \sigma} +  \frac{1}{4}\,e\, g^{\mu\rho} g^{\nu \sigma} \mathcal{R}_{\mu\nu}(T) \mathcal{R}_{\rho\sigma}(T) \Big)\label{eq:dSSugraAllOrders}\;.
\end{aligned}
\end{equation}
We have shown that,\footnote{Details of the calculation will be given in an extended version of this work.} on the surface $\Sigma\equiv\frac{\delta I}{\delta \omega_\mu{}^{ab}}=0$, the action $I[e,\bar{\psi}, \psi,A]=I\vert_\Sigma$ is invariant under the following local supersymmetry transformations,
\begin{align}
    \delta e_\mu^a &=-\frac{1}{4} (\bar{\psi}_\mu \gamma^5 \gamma^a \varepsilon_Q - \bar{\varepsilon}_Q \gamma^5 \gamma^a \psi_\mu)\;,\label{eq:gaugeSUSYTOTe}\\
    \delta \psi_\mu &=\mathcal{D}_\mu\varepsilon_Q + \frac{ i\lambda}{2 } \gamma_\mu \varepsilon_Q +\frac{ \kappa}{4\sqrt{2}} \mathcal{R}_{\nu\rho}(T) \gamma^{\nu\rho}\gamma_\mu\varepsilon_Q\;,\\
    \delta \bar{\psi}_\mu &=\mathcal{D}_\mu\bar{\varepsilon}_Q + \frac{ i\lambda}{2 } \bar{\varepsilon}_Q \gamma_\mu+\frac{ \kappa}{4\sqrt{2}}  \mathcal{R}_{\nu\rho}(T)\bar{\varepsilon}_Q\gamma_\mu\gamma^{\nu\rho}\;,\\
    \delta A_\mu &= \frac{1}{2\sqrt{2}\kappa } (\bar{\varepsilon}_Q \gamma^5 \psi_\mu - \bar{\psi}_\mu \gamma^5 \varepsilon_Q) \;,\label{eq:gaugeSUSYTOTA}
\end{align}
to all orders in the fermions, where 
$$\mathcal{D_\mu}\varepsilon_Q:=\left(\partial_\mu 
+\frac{1}{4}\omega_\mu{}^{ab}(e,\bar{\psi},\psi)\,\gamma_{ab}
-i\frac{\lambda \kappa}{\sqrt{2}}A_\mu \right)\varepsilon_Q\;.$$ 
The action above is the simplest de Sitter supergravity theory that can be constructed, and, to the best of our knowledge, it has never been given to all orders in the fermions before. It defines $\mathcal{N}=2$ de Sitter pure supergravity theory.
To show its uniqueness, we rewrite it as follows, dropping 
the terms quartic in the fermions:
\begin{equation}
\begin{aligned}
    I[e,\bar{\psi}, \psi,A] =  \int d^4x \,e\,\Bigg(&\frac{1}{2 \kappa^2} (R-2\Lambda) - \frac{1}{2 \kappa^2}\bar{\psi}_\mu \gamma^{5}\gamma^{\mu \nu \rho } \mathcal{D}_\nu\psi_\rho \\
     &+\frac{i \lambda}{2 \kappa^2} \bar{\psi}_\mu \gamma^5 \gamma^{\mu \nu} \psi_\nu+\frac{1}{4} F_{\mu \nu} F^{\mu \nu} \\
    &-\frac{1}{2\sqrt{2} \kappa}  \bar{\psi}_\mu\gamma^5(  F^{\mu \nu}-i\gamma^5 (* F)^{\mu\nu}) \psi_{\nu} \Bigg)+ \mathcal{O}(\psi^4)\,, \label{eq:ActionUptoQuarticfermions}
\end{aligned}
\end{equation}
where $\Lambda=3\lambda^2$ is the cosmological constant, 
$\mathcal{D}_\mu \psi_\nu =(\partial_\mu +\frac{1}{4}\omega_\mu{}^{ab}(e,\bar{\psi}, \psi)\, \gamma_{ab}-i\frac{\lambda \kappa}{\sqrt{2}}A_\mu)\psi_\nu$\,, and $(* F)^{\mu\nu}=\frac{1}{2}e^{-1}\varepsilon^{\mu\nu\rho\sigma}F_{\rho\sigma}$. 
Now, it is easy to see that all the cubic terms appearing above 
exactly produce the exhaustive list of cubic couplings we 
have classified in the previous section. Following standard arguments, 
explained in e.g. the conclusion of Section 3 of \cite{Boulanger:2018fei}, 
we conclude the uniqueness of the theory.

We have checked that, up to terms quartic in fermions and upon performing a 
similarity transformation of the gamma matrices, the action in eq. 
\eqref{eq:ActionUptoQuarticfermions} 
coincides with the action written down in \cite{Pilch:1984aw}. 
The important difference between our result and \cite{Pilch:1984aw} is that 
we have explicitly constructed the terms quartic in fermions, and have 
shown gauge invariance of the action to all orders.  
As expected, the spin-1 field in \eqref{eq:ActionUptoQuarticfermions} 
has the wrong sign in front of its kinetic term, making the appearance of 
negative-norm states unavoidable. However, as mentioned earlier, 
the graviphoton is not the only source of non-unitarity of the theory; 
the gravitino sector of the theory also contains negative-norm states.

\section{Conclusion and Outlook}

In this paper, we provided the complete action for $\mathcal{N}=2$ pure supergravity with positive cosmological constant and proved its uniqueness, as well as clarified that both the graviphoton and the gravitino give rise to negative-norm states from a quantum field theory perspective. We have thus extended the results of \cite{Pilch:1984aw} in several directions.
The complete action  we found in this paper opens new directions in view of the 
renewed interest and new results concerning the Euclidean approach to de Sitter 
quantum gravity \cite{Anninos:sphere, Maldacena:2024spf, Anninos:2025mje}. In 
particular, it is tempting to study the consistency of de Sitter supergravity in 
Euclidean signature, taking as our guide the sphere path integral 
\cite{Anninos:sphere, Maldacena:2024spf, Anninos:2025mje}. 
Indeed, certain reality properties that are considered problematic in the 
Lorentzian setting, such as a complex action, turn out to be harmless in the 
Euclidean theory \cite{Anninos:2025mje}.
Quantum field theory in Lorentzian signature requires a real action for quantum 
mechanical consistency of the theory, but 
the real action for the gravitino \eqref{eq:ActionS0} gives rise to an indefinite inner 
product \cite{Higuchi:2025pbc}. However, this is not necessarily problematic as,  ultimately, the gravitino should be viewed as part of gravitational system (gravitino coupled to gravity around dS$_{4}$), and the unitarity of a quantum gravitational
theory may be approached from a different perspective. Adopting a different perspective for unitarity should not come as a surprise considering the fact that de Sitter symmetries must annihilate all states in the Hilbert space already at the level of linearised gravity \cite{Higuchi:1991tk}. 
Having mentioned different perspectives, it was recently shown \cite{Anninos:2025mje} that in the Euclidean path integral approach 
it is consistent for the gravitino action  to be complex \cite{Anninos:2025mje}, 
see also \cite{Stone_2022}. 
In particular, a complex action for the gravitino on the four-sphere (Euclidean de Sitter)  leads to 1-loop 
expressions for the sphere path integral that are not 
only real-valued but also  are expressed in terms of \emph{unitary} $so(4,1)$ 
characters \cite{Anninos:2025mje}, avoiding the problems of the Lorentzian theory. 
Therefore, it appears very promising to take 
the complete action of $\mathcal{N}=2$ de Sitter 
pure supergravity that we have established in this paper, find its counterpart in Euclidean signature and study its partition function, as was done in the 2D case \cite{Anninos:2023exn}. This 
 is something we leave for future work, 
as well as the non-linear completion of the free unitary chiral 
supergravity multiplet found in \cite{Higuchi:2025pbc}.
Finally, our results concerning the explicit knowledge of the complete set 
of interaction vertices of the simplest, pure de Sitter supergravity, 
may also be useful for the determination of 
higher-point cosmological correlators for fermions, see \cite{Chen:2025foq} 
for a related recent work.

Addressing these questions will contribute to the ongoing effort to obtain a  
theoretical dataset for quantities central to (Euclidean) quantum gravity with 
positive cosmological constant.


\section*{Acknowledgments}
We would like to thank Atsushi Higuchi for many useful discussions and 
for sharing his helpful notes. We would also like to thank Dionysios Anninos,  
Evgeny Skvortsov and Philippe Spindel for useful discussions. 
The packages xTras \cite{Nutma:2013zea} and FieldsX \cite{Frob:2020gdh} 
of the xAct suite
of Mathematica packages for tensor computer algebra were used in several 
computations. S.T. is a research fellow of the F.R.S.-FNRS supported by the ASP 
fellowship FC 54793 MassHighSpin. The work of V. A. L.  is supported by the 
ULYSSE Incentive Grant for Mobility in Scientific Research [MISU] F.6003.24, 
F.R.S.-FNRS, Belgium. V. A. L. thanks the Department of Mathematics of the 
University of York (UK) for hospitality where part of his work was carried out. 
The work of N.~B. was partly supported by 
the F.R.S.-FNRS PDR grant number T.0047.24.

\appendix

\section{Gravitational coupling of the Dirac spin-3/2 field}
\label{App: coupling spin-3/2 with spin-2}
The non-Abelian cubic interaction term involving a the massless spin-2 field $h_{\mu\nu}$ and a massless Dirac spin-3/2 field, $\phi_\mu$, is $S_1^{(\phi-h)}=\int d^4x \sqrt{-\bar{g}}\, \alpha_1 L_1^{(\phi-h)}$, with
\begin{equation}
\begin{aligned}
    L_1^{(\phi-h)} = & \;\bar{\phi}_\alpha \gamma^{\alpha\beta \gamma} \gamma^{\mu \nu}\gamma^5 \phi_{ \gamma} \nabla_\mu h_{\nu\beta} 
    +2 \bar{\phi}_\delta \gamma^{\gamma \delta \alpha} \gamma^5 \nabla_\gamma \phi_{\alpha}  h \\
    &+2 \bar{\phi}_\alpha \gamma^{\gamma \delta \beta}\gamma^5 \nabla_\delta \phi_{\beta} h^{\alpha}{}_{\gamma} +2 \bar{\phi}_\delta \gamma^{\gamma\delta \alpha} \gamma^5\nabla_\alpha \phi_{\beta} h^{\beta}{}_\gamma \\
    &-2\bar{\phi}_\delta \gamma^{\gamma \delta \alpha} \gamma^5 \nabla_\beta \phi_{\alpha} h^{\beta}{}_\gamma -2 i \lambda \bar{\phi}_\beta \gamma^{\alpha \beta}\gamma^5 \phi_\gamma h_\alpha{}^\gamma\\ 
    &+2 i \lambda \bar{\phi}_\gamma \gamma^{\alpha \beta}\gamma^5 \phi_\beta h_\alpha{}^\gamma -2 i \lambda \bar{\phi}_\alpha \gamma^{\alpha \beta} \gamma^5 \phi_\beta h \;. \label{eq:VertexLGrav}
\end{aligned}
\end{equation}
The corresponding deformations of the gauge transformations are 
\begin{align}
    \delta^{(\phi-h)}_1 h_{\mu \nu} &= \alpha_1 \Bigl(\bar{\epsilon}\gamma_{(\mu} \gamma_5 \phi_{\nu)}-\bar{\phi}_{(\mu} \gamma_{\nu)}\gamma_5 \epsilon \Bigr)\,, \label{eq:DefL1sugrah}\\
    \delta^{(\phi-h)}_1 \phi_{\mu} &= \alpha_1 \Bigl(  \gamma^{\alpha \beta}\epsilon \,\nabla_\alpha h_{\beta \mu} -i \lambda \gamma^\nu \epsilon \,h_{\mu \nu} 
    + 2 i  \lambda \gamma_\nu \phi_\mu \xi^\nu - \gamma_{\alpha \beta} \phi_\mu \nabla^\alpha \xi^\beta +8  \Phi_{\mu \nu}\xi^\nu \Bigr)\,,\notag
\end{align}
$\delta^{(\phi-h)}_1 \bar{\phi}_{\mu}$ equals the Dirac conjugate of $\delta^{(\phi-h)}_1 \phi_{\mu} $\,, and $\delta^{(\phi-h)}_1 A_\mu=0$\,.
\section{Electromagnetic coupling of the Dirac spin-3/2 field}\label{App: coupling spin-3/2 with spin-1}

The non-Abelian coupling between the massless spin-1 field and the Dirac spin-3/2 is $S^{(\phi-A)}_1=\int d^4x \sqrt{-\bar{g}}\,\alpha_2 L^{(\phi-A)}_1$\,, with 
\begin{equation}
\begin{aligned}
    L^{(\phi-A)}_1=& \, 2\Bigl(\bar{\phi}_{\mu}\gamma^5(F^{ \mu\nu} -i\gamma^5(*F)^{\mu\nu})\phi_{\nu }+i \lambda A_\alpha \bar{\phi}_{ \gamma} \gamma^{\alpha \beta\gamma}\gamma^5\phi_{\beta} \Bigr)\;. \label{eq:VertexL1EM}    
\end{aligned}
\end{equation}
The corresponding deformations to the gauge transformations are
\begin{align}
    \delta^{(\phi-A)}_1 \phi_{\mu} &= 
    \alpha_2  \Bigl( - 2 i \lambda \phi_\mu \pi + 2 i \lambda A_\mu \epsilon- \frac{1}{2}F_{\alpha \beta} \gamma^{\alpha \beta} \gamma_\mu \epsilon\Bigr)\,, \notag\\
    \delta^{(\phi-A)}_1 A_\mu &= 2\,\alpha_2 \, \sigma  \Bigl(\bar{\epsilon} \gamma^5 \phi_\mu - \bar{\phi}_\mu \gamma^5 \epsilon\Bigr)\,,\label{eq:DefL1EMA}
\end{align}
$\delta^{(\phi-A)}_1 \bar{\phi}_{\mu}$ equals the Dirac conjugate of $\delta^{(\phi-A)}_1 \phi_{\mu} $\,, and $\delta^{(\phi-A)}_1 h_{\mu\nu}=0$\,.



\providecommand{\href}[2]{#2}\begingroup\raggedright\endgroup

\end{document}